\documentclass{ws-ijmpd}

\begin{document}

\title{ANALYSIS OF UNBALANCED BLACK RING SOLUTIONS WITHIN THE QUASILOCAL FORMALISM}

\author{Liu Zhen-Xing}

\address{Wuhan Institute of Physics and Mathematics, Chinese Academy of Sciences, \\
West No.30 Xiao Hong Shan, Wuhan 430071, China\\
liunenu@tom.com}

\author{Chen Ze-Qian}

\address{Wuhan Institute of Physics and Mathematics, Chinese Academy of Sciences, \\
West No.30 Xiao Hong Shan, Wuhan 430071, China\\
zqchen@wipm.ac.cn}

\maketitle

\begin{abstract}
We investigate the properties of rotating asymptotically flat black ring
solutions in five-dimensional Einstein-Maxwell-dilaton gravity with the
Kaluza-Klein coupling. Within the quasilocal formalism, the balance condition
for these solutions is derived by using the conservation of the renormalized
boundary stress-energy tensor, which is a new method proposed by Dumitru
Astefanesei and his collaborators. We also study the thermodynamics of
unbalanced black rings. The conserved charges and the thermodynamical
quantities are computed. Due to the existence of a conical singularity
in the boundary, these quantities differ from the original regular ones.
It is shown that the Smarr relation and the quantum statistical relation
are still satisfied. However, we get an extra term in the first law of
thermodynamics. As the balance condition is imposed this extra term vanishes.
\end{abstract}


\section{Introduction}	

The uniqueness theorems valid in four dimensions tell us the simplicity
of black holes. That is, an asymptotically flat, stationary black hole
solution of Einstein-Maxwell theory is completely characterized by its
mass, angular momentum and charge, and the spherical topology is the
only allowed horizon topology. However, all these conventional notions
do not apply in the framework of higher-dimensional physics. Since the
discovery of a rotating black ring solution in five dimensions by Emparan
and Reall,\cite{1} many new properties not shared by four-dimensional
black holes were found and studied. A black ring is an object equipped
with an event horizon of topology $S^1\times S^2$ rather than the much
more familiar $S^3$ topology. It was also explicitly proved in Ref. 1
the breakdown of the uniqueness in five dimensions: for vacuum solutions,
there exist one black hole and two black rings all with the same values
of the mass and the angular momentum. The spectrum of black objects
contains both black holes and black rings in five dimensions, which is
far richer than in four dimensions. This is why the uniqueness is so
weak in higher-dimensional gravity.

Many examples of black ring solutions have been constructed in various
gravity theories by now. An important question raised in studying these
solutions refers to the dynamic balance condition. The balance condition
is a constraint on the parameters of a black ring which can be understood
on physical grounds: the angular momentum of the ring must be tuned so that
the centrifugal force balances the tension and gravitational self-attraction.
In some previous literature\cite{2}\cdash\cite{4} this condition for the
thin rings was obtained by demanding the absence of all conical singularities.
Recently, Astefanesei, Rodriguez and Theisen came up with a new method
for dealing with the general (thin or fat) rings.\cite{5} They obtained
the balance condition for a vacuum solution by considering the conservation
of the renormalized boundary stress-energy tensor. This closes a gap
left unanswered in black ring physics. The main goal of the present paper
is to apply the new method proposed in Ref. 5 to analyzing the charged
dilatonic black ring solutions. We will give a systematical computation
showing how to derive the dynamic balance condition.

Employing the quasilocal formalism supplemented with boundary counterterms,
we will also carry out a preliminary study of the thermodynamics of the
unbalanced rings. The Smarr relation and the quantum statistical relation
will be checked. In particular, we are able to prove that the balance
condition is equivalent with the satisfaction of the first law of thermodynamics.

The remainder of this paper is organized as follows. In the next section we
introduce the rotating black ring solutions of the Einstein-Maxwell-dilaton
(EMd) theory with a coupling constant $\alpha=\sqrt{8/3}$. In section 3 we
give a brief review of the quasilocal formalism supplemented with the counterterms.
In section 4 we compute the divergence-free boundary stress tensor in detail
and derive the balance condition. Section 5 investigates the thermodynamics
preliminarily. Finally, we conclude in section 6 with a discussion of
our results.

\section{Charged Dilatonic Black Rings}

We consider the five-dimensional EMd system. The field equations are given by
\begin{equation}
R_{\mu\nu}=2\partial_\mu\Phi\partial_\nu\Phi+2e^{-2\alpha\Phi}\Big(F_{\mu\rho}F_{\nu}^{\ \rho}-\frac{1}{6}g_{\mu\nu}F_{\rho\sigma}F^{\rho\sigma}\Big),
\end{equation}
\vspace{-6.3mm}
\begin{equation}
\nabla_\mu\big(e^{-2\alpha\Phi}F^{\mu\nu}\big)=0,
\end{equation}
\vspace{-6.3mm}
\begin{equation}
\nabla_\mu\nabla^{\mu}\Phi+\frac{\alpha}{2}e^{-2\alpha\Phi}F_{\rho\sigma}F^{\rho\sigma}=0,
\end{equation}
where $\alpha=\sqrt{8/3}$ is the Kaluza-Klein (KK) coupling constant. A family of
rotating solutions to this theory were presented in Ref. 6:
\begin{eqnarray}
ds^2&=&-\frac{F(x)}{F(y)}\frac{\Big(dt+\big(R\sqrt{\lambda\nu}\cosh\beta\big)(1+y)d\psi\Big)^2}{V_\beta(x,y)^{2/3}}+\frac{R^2}{(x-y)^2}V_\beta(x,y)^{1/3}\nonumber\\
    &\times&\bigg[-F(x)\bigg(G(y)d\psi^2+\frac{F(y)}{G(y)}dy^2\bigg)+F(y)^2\bigg(\frac{dx^2}{G(x)}+\frac{G(x)}{F(x)}d\phi^2\bigg)\bigg],
\end{eqnarray}
\begin{equation}
A=\frac{1}{2}\frac{\sinh\beta}{\cosh^2\!\beta F(y)\!-\!\sinh^2\!\beta F(x)}\bigg[\cosh\beta \Big(F(y)-F(x)\Big)dt\!-\!R\sqrt{\lambda\nu}(1+y)F(x)d\psi\bigg],
\end{equation}
\begin{equation}
e^{-\Phi}=V_\beta(x,y)^{\scriptstyle \frac{1}{\sqrt{6}}},
\end{equation}
where
\begin{equation}
F(\xi)=1-\lambda\xi,\qquad G(\xi)=(1-\xi^2)(1-\nu\xi),
\end{equation}
and
\begin{equation}
V_\beta(x,y)=\cosh^2\!\beta-\sinh^2\!\beta\frac{F(x)}{F(y)}.
\end{equation}
$R$, $\lambda$, $\nu$ and $\beta$ are parameters whose appropriate combinations
give various physical quantities. The parameters $\lambda$ and $\nu$ lie in the
range $0\leqslant\nu<\lambda\leqslant1$. To obtain a Lorentzian signature, the
coordinates $x$ and $y$ must take values in
\begin{equation}
-1\leqslant x\leqslant1,\qquad -\infty<y\leqslant-1,\qquad \lambda^{-1}<y<\infty.
\end{equation}
Asymptotic spatial infinity is reached as $x\rightarrow y \rightarrow -1$.

We should emphasize that these solutions are not physically regular. In order to
eliminate the conical singularities at $x=-1$ and $y=-1$ we identify the angles
$\phi$ and $\psi$ with equal periods
\begin{equation}
\Delta\phi=\Delta\psi=\frac{4\pi\sqrt{F(-1)}}{\mid G^\prime (-1)\mid}=2\pi\frac{\sqrt{1+\lambda}}{1+\nu}.
\end{equation}
Then the orbits of $\partial_\phi$ and $\partial_\psi$ close off smoothly at
$x=-1$ and $y=-1$, respectively.

There are now two cases depending on the value of $\lambda$. One of them
corresponds to the black rings and the other to the black holes.

Case one is defined by $\lambda<1$. In this case, there will also be a
conical singularity at $x=+1$ unless $\phi$ is identified with period
\begin{equation}
\Delta\phi^\prime=\frac{4\pi\sqrt{F(+1)}}{\mid G^\prime(+1)\mid}=2\pi\frac{\sqrt{1-\lambda}}{1-\nu}.
\end{equation}
Demanding $\Delta\phi=\Delta\phi^\prime$ yields
\begin{equation}
\lambda=\frac{2\nu}{1+\nu^2}\qquad \mbox{(black ring)},
\end{equation}
which makes the circular orbits of $\partial_\phi$ close off smoothly also
at $x=+1$. Then $(x,\phi)$ parametrize a two-sphere $S^2$, $\psi$ parametrizes
a circle $S^1$, and the sections at constant $t$, $y$ have the topology
of a ring $S^1\times S^2$. Physically, (12) is interpreted as the balance
condition for a black ring. We will rederive this condition from a new
perspective in section 4.

Case two is defined by
\begin{equation}
\lambda=1\qquad \mbox{(black hole)}.
\end{equation}
In this case, the orbits of $\partial_\phi$ do not close at $x=+1$. Then $(x,\phi,\psi)$
parametrize a three-sphere $S^3$ at constant $t$, $y$. The line element (4) describes
the charged dilatonic black holes. To see this, it is convenient to introduce the
transformation of coordinates
\begin{eqnarray}
x&=&-1+\frac{2m\cos^2\!\theta}{r^2+a^2\cos^2\!\theta},\\
y&=&-1-\frac{2m\sin^2\!\theta}{r^2-m+a^2\cos^2\!\theta},\\
\psi&=&\frac{\sqrt{2}}{1+\nu}\tilde{\psi},\\
\phi&=&\frac{\sqrt{2}}{1+\nu}\tilde{\phi},
\end{eqnarray}
with
\begin{equation}
m=\frac{4R^2}{1+\nu},\qquad a=\frac{R\sqrt{8\nu}}{1+\nu}.
\end{equation}
In these new coordinates, the solutions take the form
\begin{eqnarray}
ds^2&=&\Big(1+\frac{m}{\Sigma_1}\sinh^2\!\beta\Big)^{-2/3}\bigg[-dt^2+\frac{m}{\Sigma_1}\Big(dt+a\sin^2\!\theta\cosh\!\beta\, d\tilde{\psi}\Big)^2\nonumber\\
&-&\frac{m}{\Sigma_1-m}\Big(a^2\sin^4\!\theta\cosh^2\!\beta\Big)d\tilde{\psi}^2\bigg]+\Big(1+\frac{m}{\Sigma_1}\sinh^2\!\beta\Big)^{1/3}\nonumber\\
&\times&\bigg[\Sigma_1\Big(\frac{1}{\Sigma_2}dr^2+d\theta^2\Big)+r^2\cos^2\!\theta\, d\tilde{\phi}^2+\frac{\Sigma_1\Sigma_2}{\Sigma_1-m}\sin^2\!\theta \,d\tilde{\psi}^2\bigg],\\
A&=&\frac{1}{2}\frac{m\sinh\!\beta}{\Sigma_1+m\sinh^2\!\beta}\bigg(\cosh\beta\,dt+a\sin^2\!\theta\,d\tilde{\psi}\bigg),\\
e^{-\Phi}&=&\bigg(1+\frac{m}{\Sigma_1}\sinh^2\!\beta\bigg)^{\frac{1}{\sqrt{6}}},
\end{eqnarray}
where
\begin{equation}
\Sigma_1\equiv r^2+a^2\cos^2\!\theta,\qquad \Sigma_2\equiv r^2+a^2-m.
\end{equation}
If we set $\beta=0$, the neutral Myers-Perry black hole solutions with
only one non-vanishing angular momentum (see Ref. 7) will be recovered.

Both for black rings and black holes, $|y|=\infty$ is an ergosurface,
$y=1/\nu$ is the event horizon, and the inner, spacelike singularity is reached
as $y\rightarrow\lambda^{-1}$ from above.

We will focus on the black rings $(0<\lambda<1)$ in this paper.

\section{Quasilocal Formalism}
It is well known that the concept of local energy (defining the energy
at a spacetime point) in general relativity is meaningless. However,
one can define a quasilocal energy in a spatially bounded region by
employing the quasilocal formalism of Brown and York.\cite{8} For
asymptotically flat spacetimes, the quasilocal energy agrees with
the ADM energy\cite{9} in the limit that the boundary tends to
spatial infinity.

The quasilocal formalism provides a powerful method to compute the
conserved charges and to study the thermodynamics of black objects.
Within this formalism, the conserved charges are related to the
divergence-free boundary stress tensor:\cite{10}
\begin{equation}
\tau_{ij}\equiv\frac{-2}{\sqrt{-h}}\frac{\delta I}{\delta h^{ij}}=\frac{1}{8\pi}\Big(K_{ij}-h_{ij}K-\Psi\big(\mathcal{R}_{ij}-\mathcal{R}h_{ij}\big)
-h_{ij}\Box\Psi+\Psi_{;ij}\Big),
\end{equation}
where $\mathcal{R}_{ij}$ is the Ricci tensor of the induced boundary metric
$h_{ij}$, $\Psi\equiv\sqrt{\frac{3}{2\mathcal{R}}}$, $K_{ij}$ is the extrinsic
curvature of the boundary, and
\begin{eqnarray}
I=I_{B}+I_{\partial B}=&-&\frac{1}{16\pi}\int_B \Big(R-2g^{\mu\nu}\partial_\mu\Phi\partial_\nu\Phi-e^{-2\alpha\Phi}F_{\mu\nu}F^{\mu\nu}\Big)\sqrt{-g}\,d^5x\nonumber\\
&-&\frac{1}{8\pi}\int_{\partial B}\Bigg(K-\sqrt{\frac{3}{2}\mathcal{R}}\Bigg)\sqrt{-h}\,d^4x
\end{eqnarray}
is the total action which has been renormalized by the counterterm\cite{11}
\begin{equation}
I_{ct}=\frac{1}{8\pi}\int_{\partial B}\sqrt{\frac{3}{2}\mathcal{R}}\sqrt{-h}\,d^4x.
\end{equation}
We use Greek indices and Latin indices to denote the bulk coordinates and the
boundary coordinates, respectively.

The boundary metric can be written as
\begin{equation}
h_{ij}dx^idx^j=-N^2dt^2+\sigma_{ab}(dy^a+N^adt)(dy^b+N^bdt),
\end{equation}
where $N$ is the lapse function, $N^a$ is the shift vector, and $\{y^a\}$ are
the intrinsic coordinates on a closed hypersurface $\Sigma$ (with normal $\tilde{n}^i$).
Then the mass and the angular momentum are defined by
\begin{eqnarray}
M&=&-\oint_\Sigma d^3y\sqrt{\sigma}\,\tilde{n}^i\tau_{ij}\xi^j_t,\\
J_\psi&=&-\oint_\Sigma d^3y\sqrt{\sigma}\,\tilde{n}^i\tau_{ij}\xi^j_\psi,
\end{eqnarray}
where $\xi_t=\partial_t$ and $\xi_\psi=\partial_\psi$ are normalized Killing vectors.

Note that the boundary stress tensor satisfies an approximate local conservation law\cite{8}
\begin{equation}
D^i\tau_{ij}=-n^\mu T_{\mu j}\equiv-T_{nj},
\end{equation}
where $D^i$ is the covariant derivative of $h_{ij}$, $n^\mu$ is the normal on
the boundary, and $T_{\mu\nu}$ is the energy-momentum tensor. We will use this
conservation law to derive the balance condition for a charged black ring in
the next section.

\section{Balance Condition}

We consider the black rings, i.e., $\lambda\neq1$. The aim of this section is
to rederive the balance condition (12) from the conservation law (29). As done
in Ref. 5, we just remove the conical singularity in the bulk by identifying
$\phi$ and $\psi$ with an equal period
\vspace{-3.5mm}
\begin{equation}
\Delta\phi=\Delta\psi=\frac{4\pi\sqrt{F(+1)}}{\mid G^\prime(+1)\mid}=2\pi\frac{\sqrt{1-\lambda}}{1-\nu}.
\end{equation}
The conical singularity in the boundary $x=y=-1$ still exists. We introduce
the transformation of coordinates
\begin{eqnarray}
x&=&-1+\frac{2m\cos^2\!\theta}{r^2+a^2\cos^2\!\theta},\\
y&=&-1-\frac{2m\sin^2\!\theta}{r^2-m+a^2\cos^2\!\theta},\\
\psi&=&\frac{\sqrt{1-\lambda}}{1-\nu}\tilde{\psi},\\
\phi&=&\frac{\sqrt{1-\lambda}}{1-\nu}\tilde{\phi},
\end{eqnarray}
with
\begin{equation}
m=\frac{(1+\lambda)^2R^2}{1+\nu},\qquad a=\frac{R\sqrt{(1+\lambda)(\lambda-1+\nu+3\lambda\nu)}}{1+\nu}.
\end{equation}
In these new coordinates, the asymptotic form of the metric is
\begin{eqnarray}
g_{tt}&=&-1+\frac{2}{3}\frac{(1+2\cosh^2\!\beta)R^2\lambda(1+\lambda)}{1+\nu}\frac{1}{r^2}+O(1/r^4),\\
g_{t\psi}&=&\frac{2R^3(1+\lambda)^2\sqrt{\lambda\nu(1-\lambda)}\sin^2\!\theta\cosh\beta}{1-\nu^2}\frac{1}{r^2}+O(1/r^4),\\
g_{rr}&=&1+\frac{2}{3}R^2\frac{(1+\lambda)}{(1+\nu)^2}\Big[3(2\nu-\lambda+1)\cos^2\!\theta+\lambda(1+\nu)\cosh^2\!\beta\nonumber\\
&+&(2\lambda-\lambda\nu-3\nu)\Big]\frac{1}{r^2}+O(1/r^4),\\
g_{\theta\theta}&=&r^2+\frac{2}{3}R^2\frac{(1+\lambda)}{(1+\nu)^2}\Big[3(2\nu-\lambda+1)\cos^2\!\theta\nonumber\\
&+&(1+\nu)(\lambda\cosh^2\!\beta+2\lambda-3)\Big]+O(1/r^2),\\
g_{r\theta}&=&-\frac{R^4(1-\nu)(1-\lambda)(1+\lambda)^3\sin4\theta}{4(1+\nu)^3}\frac{1}{r^3}+O(1/r^5),\\
g_{\psi\psi}&=&\frac{(1-\lambda)(1+\nu)^2}{(1-\nu)^2(1+\lambda)}r^2\sin^2\!\theta+\frac{2}{3}R^2\frac{(1-\lambda)}{(1-\nu)^2}\Big[
(1+\nu)\lambda\cosh^2\!\beta\nonumber\\
&+&(2\nu-1)\lambda+3\nu\Big]\sin^2\!\theta+O(1/r^2),\\
g_{\phi\phi}&=&\frac{(1-\lambda)(1+\nu)^2}{(1-\nu)^2(1+\lambda)}r^2\cos^2\!\theta+\frac{2}{3}R^2\frac{(1-\lambda)}{(1-\nu)^2}
(1+\nu)(\lambda\cosh^2\!\beta\nonumber\\
&+&2\lambda-3)\cos^2\!\theta+O(1/r^2).
\end{eqnarray}
Note that we still use $\psi$ and $\phi$ to denote the angles after performing
the transformation. The non-vanishing stress tensor components are
\vspace{1mm}
\begin{eqnarray}
\tau_{tt}&=&\frac{1}{8\pi}\bigg(-R^2\frac{(1+\lambda)}{(1+\nu)^2}\Big[\lambda(1+\nu)(1+2\cosh^2\!\beta)\nonumber\\
&+&\frac{10}{3}(2\nu-\lambda+1)\cos2\theta\Big]\frac{1}{r^3}+O(1/r^5)\bigg),\\
\tau_{t\psi}=\tau_{\psi t}&=&\frac{1}{8\pi}\bigg(-\frac{4R^3(1+\lambda)^2\sqrt{\lambda\nu(1-\lambda)}\sin^2\!\theta\cosh\beta}{1-\nu^2}\frac{1}{r^3}+
O(1/r^5)\bigg),\\
\tau_{\theta\theta}&=&\frac{1}{8\pi}\bigg(\frac{4}{3}R^2\frac{(1+\lambda)}{(1+\nu)^2}(2\nu-\lambda+1)\cos2\theta\frac{1}{r}+O(1/r^3)\bigg),\\
\tau_{\psi\psi}&=&\frac{1}{8\pi}\bigg(\frac{4}{3}R^2\frac{(1-\lambda)}{(1-\nu)^2}(2\nu-\lambda+1)
(-1+2\cos2\theta)\sin^2\!\theta\frac{1}{r}+O(1/r^3)\bigg),\nonumber\\
&&\\
\tau_{\phi\phi}&=&\frac{1}{8\pi}\bigg(\frac{4}{3}R^2\frac{(1-\lambda)}{(1-\nu)^2}(2\nu-\lambda+1)
(1+2\cos2\theta)\cos^2\!\theta\frac{1}{r}+O(1/r^3)\bigg).\nonumber\\
\end{eqnarray}
We also compute $T_{nj}$ and find that
\begin{equation}
T_{nt}=T_{n\psi}=T_{n\phi}=0,
\end{equation}
\begin{equation}
T_{n\theta}=\frac{1}{8\pi}\Bigg[4R^6\frac{\lambda\nu(1+\lambda)^4}{(1+\nu)^4}\bigg(\frac{2}{3}\lambda\cosh^2\!\beta+\frac{\lambda}{3}-1\bigg)\sinh^2\!\beta\sin2\theta\frac{1}{r^7}
+O(1/r^9)\Bigg].
\end{equation}

In order to obtain the balance condition, we use the conservation law
\begin{eqnarray}
D^i\tau_{it}&=&0,\\
D^i\tau_{i\theta}&=&-T_{n\theta},\\
D^i\tau_{i\psi}&=&0,\\
D^i\tau_{i\phi}&=&0,
\end{eqnarray}
where $D^i$ corresponds to the regular boundary metric (induced from the
five-dimensional flat metric)
\begin{equation}
ds^2=-dt^2+r^2\big(d\theta^2+\sin^2\!\theta d\psi^2+\cos^2\!\theta d\phi^2\big).
\end{equation}
The coordinate $r$ in (54) is a constant, which determines the location of the
boundary. One can check that (50), (52) and (53) are exactly satisfied. The only
non-trivial equation is (51), which has an explicit expression as below:
\begin{equation}
\frac{8}{3}R^2\Bigg[\frac{(1+\lambda)}{(1+\nu)^2}-\frac{(1-\lambda)}{(1-\nu)^2}\Bigg](2\nu-\lambda+1)\frac{(1-2\sin^2\!2\theta)}{\sin2\theta}\frac{1}{r^3}+O(1/r^5)\nonumber
\end{equation}
\begin{equation}
=-4R^6\,\frac{\lambda\nu(1+\lambda)^4}{(1+\nu)^4}\bigg(\frac{2}{3}\lambda\cosh^2\!\beta+\frac{\lambda}{3}-1\bigg)\sinh^2\!\beta\sin2\theta\frac{1}{r^7}
+O(1/r^9).
\end{equation}
It turns out that the source term on the right-hand side of (55) falls off
sufficiently fast at infinity. This strongly supports the assumption made
in Ref. 5. Comparing both sides of (55), we find that the coefficient of
$r^{-3}$ must be identically equal to zero, namely,
\begin{equation}
\Bigg[\frac{(1+\lambda)}{(1+\nu)^2}-\frac{(1-\lambda)}{(1-\nu)^2}\Bigg](2\nu-\lambda+1)=0.
\end{equation}
Note that $2\nu-\lambda+1\neq0$. Therefore, we obtain
\begin{equation}
\lambda=\frac{2\nu}{1+\nu^2},\nonumber
\end{equation}
which is exactly the balance condition (12) for a charged dilatonic black ring.

\section{Thermodynamics}
We will discuss the thermodynamics of unbalanced rings in this section. The
quasilocal formalism defines a renormalized action (24), which is related to
the following grand-canonical potential:
\begin{equation}
G=IT,
\end{equation}
where $T$ is the temperature. One finds that
\begin{eqnarray}
&\lim\limits_{r\rightarrow\infty}\bigg(K-\sqrt{\frac{3}{2}\mathcal{R}}\bigg)\sqrt{-h}\nonumber
=-\frac{1}{6}R^2\frac{\lambda(1-\lambda)(1+\nu)}{(1-\nu)^2}\big(1+2\cosh^2\!\beta\big)\sin2\theta+O(1/r^2)&\nonumber.\\
&&
\end{eqnarray}
The expression for the boundary action is
\begin{equation}
I_{\partial B}=\frac{\pi}{12}R^2\,\frac{\lambda(1-\lambda)(1+\nu)}{(1-\nu)^2}\big(1+2\cosh^2\!\beta\big)\frac{1}{T}.
\end{equation}
The bulk action is computed by adopting the ``quasi-Euclidean'' method of Ref. 12. We obtain
\begin{equation}
I_B=-\frac{\pi}{6}R^2\,\frac{\lambda(1-\lambda)(1+\nu)}{(1-\nu)^2}\sinh^2\!\beta\frac{1}{T}.
\end{equation}
Then
\begin{equation}
G=(I_B+I_{\partial B})\,T=\frac{\pi}{4}R^2\,\frac{\lambda(1-\lambda)(1+\nu)}{(1-\nu)^2}.
\end{equation}

We use (27) and (28) to compute the mass and the angular momentum
\begin{eqnarray}
M&=&\frac{\pi}{4}R^2\,\frac{\lambda(1-\lambda)(1+\nu)}{(1-\nu)^2}\big(1+2\cosh^2\!\beta\big),\\
J_\psi&=&\frac{\pi}{2}R^3\,\frac{(1+\nu)(1+\lambda)(1-\lambda)^{3/2}\sqrt{\lambda\nu}}{(1-\nu)^3}\cosh\beta.
\end{eqnarray}
The temperature, entropy, angular velocity, electric charge and electric potential (evaluated
on the horizon) are given by
\begin{eqnarray}
T&=&\frac{1}{4\pi R}\frac{1-\nu}{\sqrt{\lambda(\lambda-\nu)}}\frac{1}{\cosh\beta},\\
S&=&2\pi^2R^3\,\frac{(1-\lambda)(\lambda-\nu)^{3/2}\sqrt{\lambda}}{(1-\nu)^3}\cosh\beta,\\
\Omega&=&\frac{1}{R}\frac{(1-\nu)}{(1+\nu)}\frac{\sqrt{\nu}}{\sqrt{\lambda(1-\lambda)}}\frac{1}{\cosh\beta},\\
Q&=&\pi R^2\,\frac{\lambda(1-\lambda)(1+\nu)}{(1-\nu)^2}\sinh\beta\cosh\beta,\\
\phi_{_{\scriptstyle H}}\!\!&=&\frac{1}{2}\tanh\beta.
\end{eqnarray}
Note that the electric charge is defined by
\begin{equation}
Q=\frac{1}{8\pi}\oint_\infty \!e^{-2\alpha\Phi}F_{\mu\nu}d\Sigma^{\mu\nu}.
\end{equation}

One can easily verify that the Smarr relation and the quantum statistical relation
for unbalanced rings are still satisfied, i.e.,
\begin{eqnarray}
M&=&\frac{3}{2}\big(T S+\Omega J_\psi\big)+\phi_{_{\scriptstyle H}}Q,\\
G&=&M-T S-\Omega J_\psi-\phi_{_{\scriptstyle H}}Q.
\end{eqnarray}
However, we get an extra term in the first law of thermodynamics, i.e.,
\begin{equation}
dM-TdS-\Omega \,dJ_\psi-\phi_{_{\scriptstyle H}}dQ=-\frac{1}{4}\pi R^2\!\frac{1}{1-\nu}d\lambda+\frac{1}{2}\pi R^2\frac{(1+\lambda)(1-\lambda)}
{(1+\nu)(1-\nu)^2}d\nu.
\end{equation}
The extra term on the right-hand side of (72) stems from the stresses due to
the conical singularity in the boundary. In other words, the fact that the
first law of thermodynamics is not satisfied is reflected in the existence
of additional stresses that deform the boundary metric. As the balance condition
is imposed this extra term vanishes. Conversely, solving the differential equation
\begin{equation}
-\frac{1}{4}\pi R^2\!\frac{1}{1-\nu}d\lambda+\frac{1}{2}\pi R^2\frac{(1+\lambda)(1-\lambda)}
{(1+\nu)(1-\nu)^2}d\nu=0
\end{equation}
can also yield the balance condition (12). Therefore, we have proved that the dynamic
balance condition is equivalent with the satisfaction of the first law of thermodynamics.

Supplemented with counterterms, the quasilocal formalism is a very powerful tool
to study the thermodynamics of various black objects. See Refs. 10, 5, 13 and 14 for further details.

\section{Conclusions}
In this work, we derived the dynamic balance condition for a charged dilatonic
black ring in a new way: considering the conservation of the renormalized
boundary stress-energy tensor. We have also investigated the properties
of unbalanced rings within the quasilocal formalism. By computing the
conserved charges and the thermodynamical quantities, we found that although
the black ring is in a state of non-equilibrium, the Smarr relation and the
quantum statistical relation are still satisfied. The conical singularity
is reflected in an extra term occurring in the first law of thermodynamics.
Particularly, we proved that the dynamic balance condition is equivalent with
the satisfaction of the first law of thermodynamics.

The quasilocal formalism supplemented with counterterms is very robust and
it provides a complete way to study the thermodynamics of black objects.
An appealing feature of the counterterm method is that the difficulties
associated with the choice of a reference background can be avoided.
Generally speaking, an action (containing both bulk term and Gibbons-Hawking
term) can be regularized by adding counterterms (boundary terms) that depend
on the intrinsic geometry of the regularizing surface. We emphasize that
the counterterm (25) used in this paper applies only to asymptotically
flat black ring solutions. It is expected that the counterterm method
will also be developed in studying the properties of nonasymptotically
flat black rings (e.g., Ref. 15).

\section*{Acknowledgments}

We are grateful to Prof. Wu Shuang-Qing for useful discussions. This
research was supported by NSFC grant No. 10775175.


\end{document}